\pdfoutput=1

\RequirePackage{fix-cm}

\documentclass[smallextended]{svjour3}       

\smartqed  

\usepackage{graphicx}
\usepackage{amsmath,amssymb, graphics, setspace}
\usepackage[export]{adjustbox}

 \journalname{General Relativity and Gravitation }

\begin{document}

\title{Space-borne Gravitational Wave Observatories}
\subtitle{}


\author{Stefano Vitale\\ \emph{For the eLISA consortium\\ and\\ the LISA Pathfinder team}}

\authorrunning{Stefano Vitale} 

\institute{S. Vitale \at
              Department of Physics, University of Trento and INFN Trento Institute for Fundamental Physics and Application\\I-38123, Povo, Trento, Italy \\
              \email{Stefano.Vitale@unitn.it} }

\date{Received: January, 10 2014 / Accepted: April 7, 2014}

\maketitle

\begin{abstract}
The paper describes the progress toward  a space-borne gravitational wave observatory and its foreseeable science potential. In particular the paper describes the status of the LISA-like mission called eLISA, the reference mission for the Gravitational Universe theme adopted by ESA for its Large mission L3, and the status of its precursor LISA Pathfinder, due to launch in 2015.


\end{abstract}

\section{Introduction}
\label{intro}

Space-borne gravitational wave (GW) detectors are aimed at observing waves in the sub-Hz range, down to sub-mHz frequencies, a range which is not accessible from ground because of gravitational noise. Within  this frequency range, most of the types of sources we know of today are expected to be abundant and powerful. Such sources are super-massive black-hole (BH) binaries,  in-spirals of a stellar mass BH into a super-massive one, and compact binaries in our galaxy. In addition, many theories of the early universe in the TeV era, also predict a relic GW background that may have substantial power in this same frequency range.

The most mature concept for a space borne gravitational wave detector is LISA \cite{NphysB}, studied jointly, for over a decade, by NASA and ESA . Many variants of this concept  have been studied over the years, including one, eLISA  \cite{Gravitational Universe} , which  is now the basis for a large observatory mission recently inserted by ESA in their long term planning for the L3 \cite{L2L3} launch slot. Other variations of the LISA concept can be found in  \cite{Seto}, \cite{Alia}, and  \cite{SGO}.

In LISA and its derivations, the gravitational wave signal can be described  as an  effective tidal acceleration of  test-masses (TMs)  pairs \cite{Congedo}.  These relative accelerations are measured from the time-modulation, by Doppler effect, of the frequency of a laser beam exchanged between the TMs in each pair. ESA has devoted a mission, LISA Pathfinder \cite{LPF}, to the demonstration of this measurement concept. The mission is due to launch in 2015.

This paper is organised as follows. We first summarise the LISA measurement concept, and the status of its eLISA implementation. We gives a summary of the science goals of eLISA and of the expected signals, and we finally give an account of the concept and of the implementation stage of LISA Pathfinder.

\begin{figure}
\begin{center}
\includegraphics[width=0.81\textwidth]{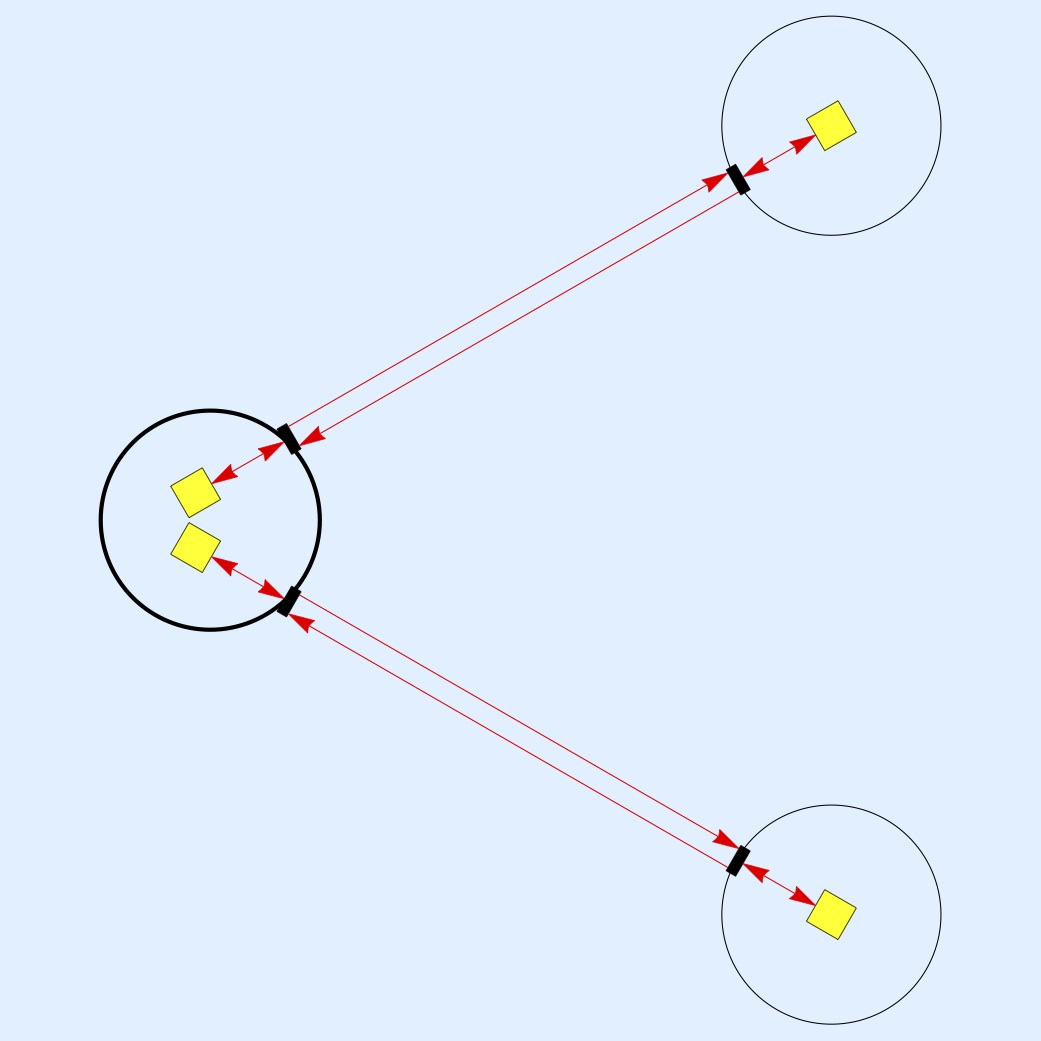}
\end{center}
\caption{Schematics of the eLISA Mother-Daughter configuration. The yellow squares represent the test-masses. The black circles indicate the satellites. The short, two-headed red lines indicate the local interferometers, while the two antiparallel arrows in between a pair of satellites   represent the two 1-million kilometre counter-propagating  links}

\label{fig:1}   

\end{figure}

\section{The eLISA detector}
\label{sec:1}

eLISA is the variation of the LISA type of mission that stems from the concept that has been  recently studied by ESA, as an ESA led mission, under the name of New Gravitational Observatory (NGO)\cite{NGO}. As said, it is the reference mission for the gravitational wave observatory to be flown by ESA as L3. Here we describe the mission as it has come out from the NGO study and  been proposed, as the reference mission, for L3 \cite{Gravitational Universe}. eLISA consists (see figure \ref{fig:1}) of three satellites, separated by one million kilometres, orbiting the sun in approximately circular orbits. With a proper choice of orbits initial anomalies and inclinations, the satellites remain approximately at the vertices of an equilateral triangle.

As in all implementation of the LISA concept, a detector "arm" is constituted by two satellites that send each other a laser beam.  An intervening GW will modulate the frequency of the received laser beam at any of the two satellites. This elementary complement, an emitter satellite, a propagating laser beam ,and a receiver satellite that detects the frequency modulation, is commonly called a link.  Each arm contains thus two counter-propagating links. The two links  in one arm are phase-locked at one end, mimicking  the  mirror reflection of a Michelson interferometer arm,  unfeasible over 1 Million kilometre, due to the intensity loss caused by  beam divergence. 

Within a link, any acceleration of either the emitting satellite or of the receiving one,  relative to their respective local inertial frame, would superimpose to the effect of GWs. These accelerations are expected to be large and noisy, mostly because of solar radiation pressure. This is why  each spacecraft carries inside, for each link terminating into it, a well shielded, nominally free falling TM that is used as a local inertial reference. The acceleration of the satellite relative to the TM, along the link direction, is measured by a local interferometer, and subtracted from  the link signal. The  corrected  signal carries then just the effect of the GW, and that of the small, residual acceleration of the TM relative to the local inertial frame\footnote{In addition to those effect, the signal would also carry the large frequency drift due to the secular change of the length of the arm. This signal is out of the measurement band of the instrument and is effectively suppressed by the data processing}. This way the link can be thought of as intervening  directly between TMs, instead of  satellites. To avoid that the satellite may collide with the inertial TM, a feedback loop, driven by the local interferometer signal, act on a set of $\mu N$ thrusters so that the satellite remain centred on the TM. This avoids the need to act with forces directly on TMs,  at least along the direction of the arm. Along some other degrees of freedom  instead, weak electrostatic forces keep the test mass steady relative to the satellite.
eLISA  carries two arms. These arms join into a central "mother" satellite while, at the other end, they terminate into two equal "daughter" satellites. The mother satellite carries then two reference TMs, while the daughters only carry one each. LISA carries three arms and then all satellites are equal and similar to the eLISA mother satellite. This has  some impact on the total mass of the mission.

eLISA satellite orbits are Earth-trailing, drift-away orbits into which the three satellites are injected, from a parking low Earth orbit, by their three individual propulsion modules. With proper adjustment of the injection, these orbit guarantee a quasi-stable formation, and  science operations then, for a bit more than 6 years. Longer operation times could be achieved by stabilising the orbits against the drift, like in LISA, by using  bigger propulsion modules with, again, some impact on the overall mission mass.

\begin{figure}
\begin{center}
  \includegraphics[width=.8\textwidth]{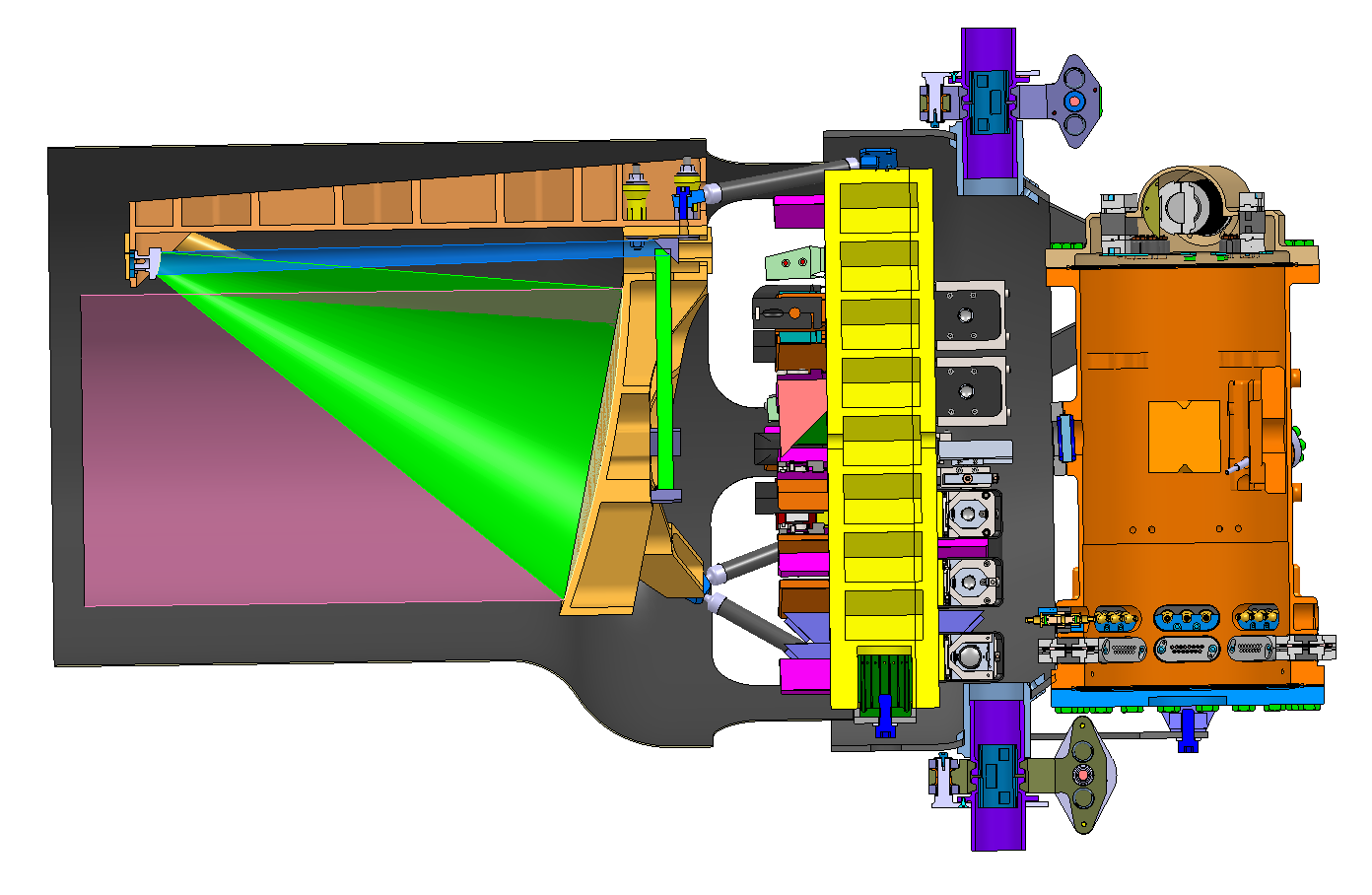}
  \end{center}

\caption{A cut of the eLISA instrument. On the left,  the off-axis telescope that exchanges the laser beams with the distant satellite. The yellow structure in the middle is a side view of the optical bench. The orange element on the right is the vacuum enclosure of the GRS in the centre of which the Au-Pt TM is also visible.(\emph{ Courtesy of Astrium GmbH, Friedrichshafen})}

\label{fig:2}   

\end{figure}

The core of the eLISA payload is the so called \textit{instrument} that terminates each link at both end. It consists of: a \textit{telescope}, that sends and receives the laser beams; an \textit{optical bench} on which all optical components reside, both for the satellite-to-satellite interferometry and for the local TM-to-satellite interferometry; finally a free, quasi-cubic, AuPT, 2 kg TM, floating inside  a housing with 3 to 4 mm clearance in all directions. The housing  carries a set of electrodes needed for motion sensing and actuation in some of the degrees of freedom. The TM , the housing and a series of devices $\--$ launch lock, TM in-orbit release mechanism, gravitational balance mass,  vacuum chamber for ground handling $\--$ form the so called \textit{gravity reference sensor} (GRS). A rendering of eLISA instrument is shown in  figure \ref{fig:2} 

\begin{figure}
\begin{center}
  \includegraphics[width=1\textwidth]{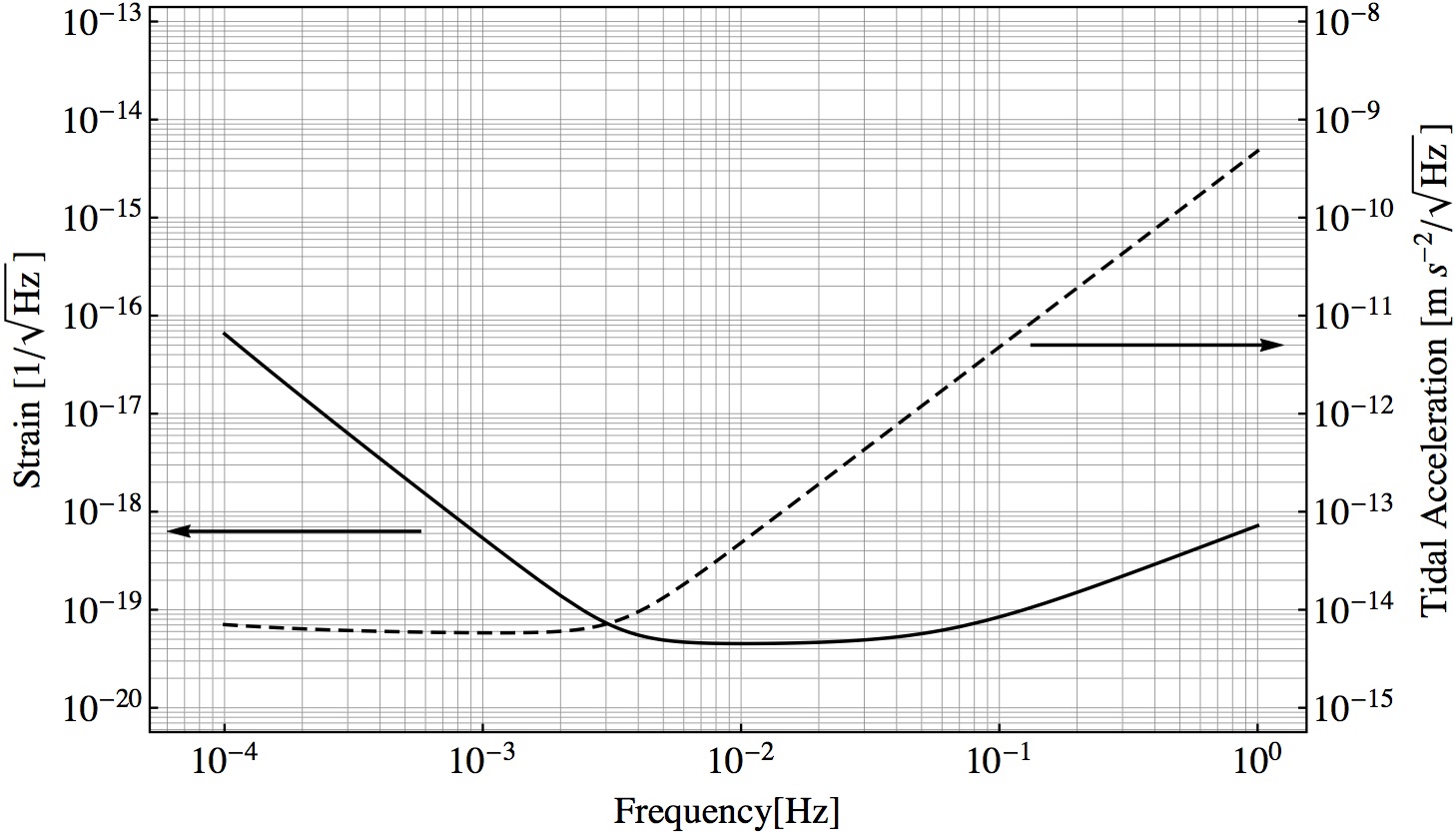}
  \end{center}

\caption{The eLISA sensitivity curve. The solid line, to be compared to the left vertical axis, is the sky-averaged sensitivity expressed as an amplitude spectral density of GW amplitude. The dashed line, to be read on the right vertical axis, is the corresponding  amplitude spectral density of the differential acceleration of the TMs at the two ends of a single link. We name this acceleration  as tidal.}

\label{fig:3}   

\end{figure}

The sensitivity curve of eLISA is shown in \ref{fig:3} , both as the conventional strain amplitude spectral density plot  and as an equivalent, single link TM tidal acceleration noise. 

\section{eLISA Science}
\label{sec:2}
A detailed discussion of eLISA science goes far beyond the scope of this progress report. Updated summaries of the large amount of work done on this revolutionary science can be found in \cite{Gravitational Universe}, \cite{NGO} and \cite{Science}. Here we  briefly summarise the observing power of eLISA, \emph{just based on our current knowledge}, and give some hint of the scientific impact that these observations are expected to have. Both discussions are heavily based on \cite{Gravitational Universe} 

\subsection{eLISA observational power}
\label{sec:2.1}

eLISA  will observe \emph{massive BH binaries} over a wide range of redshift and mass. 

Signals from these sources enter eLISA  sensitivity band at the lower end, and  drift toward higher frequencies as the in-spiral gets faster and faster, till eventually the two BHs  merge into a single one.  The resulting distorted event horizon will continue to oscillate and radiate until a rotating Kerr BH is finally formed.
The waveform for the radiated signal can be calculated, with high accuracy, by using a  Post Newtonian expansion in the early phase of the in-spiral, and by a fully general relativistic numerical simulations of black hole coalescence, in the final merger phase. With these waveforms it is possible to calculate the expected signal-to-noise ratio (SNR) for detection. In table \ref{tab:1} we report the calculated  SNR \cite{Gravitational Universe} \cite{NGO} for an equal mass, non-spinning binary,  for some special redshifts and mass values.

\begin{table}
\begin{center}
\caption{eLISA SNR for equal-mass, non spinning BH binaries}
\label{tab:1}      
\scalebox{0.85}{\begin{tabular}{llll}
\hline\noalign{\smallskip}
z &SNR$\leq$20 & Maximum SNR&Mass for maximum SNR   \\
\noalign{\smallskip}\hline\noalign{\smallskip}
 1.5 & $ 7\times10^3 M_ {\odot} \leq Mass \leq 4\times 10^8  M_ {\odot}$& $SNR >1000$&$M = 1.2\times10^6 M_ {\odot}$\\
11 & $2\times10^4 M_ {\odot} \leq Mass \leq 2\times 10^7  M_ {\odot}$& $SNR >150$&$M = 3.5\times10^5 M_ {\odot}$\\
20 & $ 3\times10^4 M_ {\odot} \leq Mass \leq 2\times 10^6 M_ {\odot}$& $SNR >50 $&$M = 1.6\times10^5 M_ {\odot}$\\
\noalign{\smallskip}\hline
\end{tabular}
}
\end{center}
\end{table}
The SNR is calculated by averaging over all possible source sky locations and wave 
polarisations, assuming two-year observations. The close z=2 range corresponds to the epoch of maximum star formation rate in the Universe and of  the highest activity of galactic nuclei.

The era around z = 11 is often called the cosmic dawn, when the universe reionized and  became transparent to stars and quasars.  The first galaxies and the first massive black holes formed at some point in the universe at $z \gtrsim 15$ or so.  As long as these first black holes are not more massive than about $10^5$ solar masses, eLISA will detect them when they form binaries out to a redshift $z \sim 20$.

Thanks to the large SNR, source parameters can be measured with high resolution. Individual (redshifted) masses can be measured with an error of $0.1 \%\--1 \% $
while the spin can be measured with  $0.01 \-- 0.1$  uncertainty. Current models predict  rates of $10\--  100$ per year. For more than 10$\%$ of the sources   the distance 
can be determined to better than a few percent and the sky location  to better than a few degrees.

eLISA will also observe the highly relativistic fall of a compact star, neutron star or BH, into a massive black hole, the so called \emph{Extreme Mass Ratio In-spiral} (EMRI).  EMRIs can 
be tracked, through the emitted GWs,  for up to $10^4 \-- 10^5 $ complex relativistic orbits close to the horizon of the central BH. eLISA will detect EMRIs with an $SNR > 20$ 
in the mass interval for the central black hole between $10^4 M_ {\odot} < M < 5 \times10^6 M_ {\odot}$ out to redshift $ z \sim 0.7$

The estimated detection rate is about 50 events for a 2 year mission, with an order of magnitude uncertainty coming from the lack of knowledge of the  actual dynamics of dense stellar nuclei .

From the observation of EMRIs one can measure: the masses of both the falling BH and that of the central massive BH with an accuracy , in most cases, better than $10^{-4}$; the eccentricity of the orbit, at the time of the fall into the event horizon, to better than $10^{-4}$; the spin of the central BH   to better than $10^{Ð3}$; the 
quadrupole moment of the central MB to better than $1\%$ for the best signals. In particular this last measurement is crucial for the detection of possible violations of the "no hair theorem" (see sect. \ref{sec:2.2})

Processes  at very high energies taking place in the 
primordial Universe can produce a \emph{stochastic background 
of gravitational waves}.  The  frequency  of the gravitational 
waves is set by the size of the horizon , i.e. by the 
temperature of the Universe at the time of the wave production $f\simeq 0.1\ mHz \left( {{{k}_{B}}T}/{1\ Tev} \right)$. 

The eLISA frequency band of 0.1 mHz to 100 mHz correspond 
to 1 to 1000 TeV, an energy scale overlapping, at its lower end, that of the Large Hadron Collider (LHC)  .
eLISA  sensitivity would allow the detection of a relic stochastic background with an energy density of 
$\Omega_{GW} \simeq 10^{-10}$ of the energy density of the Universe is converted 
to gravitational radiation at the time of production.

The majority of stars are part of binary or multiple star 
systems. A good fraction of these evolve 
into compact systems formed by  white dwarfs and/or neutron 
stars.  Stellar mass black holes binaries are also expected to exist. 
The most compact,  known as \emph{ultra-compact binaries}, have rotational frequency
in the mHz  range and are well detectable sources of GW for eLISA, if they belong to the Milky Way. 

Currently fewer than 50 ultra-compact binaries are 
known, but several of those emit strong enough gravitational waves to be detectable within weeks by eLISA, thus constituting  \emph{verification}  sources. Examples are:   HM Cnc, emitting at  $\simeq 6.2 mHz$;  V407 Vul at$ \simeq 3.7 mHz$;  and SDSS J0651+28 at $\simeq 2.8 mHz$.

In addition, several thousand binaries are expected to be resolved by eLISA with  $SNR >7$. Their rotational period,  from minutes to tens of minutes, will be measured from the frequency of the GW wave. For many binaries it will be possible to measure the chirp mass and the luminosity distance  from the  first time derivative of the wave frequency. 
For more than 100 sources concentrated around the inner Galaxy, we expect distance estimates with accuracies better than $1 \%$, allowing thus to estimate their distance to the Galactic Centre \emph{in a way that is completely independent of all other methods of comparable resolution}.

However the signals from the largest part  of ultra-compact binaries will not be separated in frequency enough to allow for individual detection. Thus they will form an 
unresolved foreground with a spectral density comparable to that of the the instrument 
noise in the mHz range, from which it could be discriminated though, thanks  to its strong yearly modulation.

\subsection{Science impact}
\label{sec:2.2}
 
The science impact of the observational power discussed in section \ref{sec:2.1} is reviewed in  \cite{Science},  and a summary is given in \cite{Gravitational Universe}. Here we want just highlight, largely quoting from that work,  how this impact spans across  astrophysics, cosmology and fundamental physics at large.

For instance  BH mergers are key markers within the evolution  the universe. As basically all BH binaries, during their evolution, irradiate a considerable power within the eLISA band, eLISA will make the largest and deepest survey of massive BH mergers. The mass sequence of mergers , from the lightest objects to the most massive ones, will  allow to distinguish between the different model for BH formation and evolution. Furthermore, the spin distribution  is expected to trace the accretion dynamics in the different stages of the merger chain.

However massive black-hole mergers are also a test bed for gravity in the non linear regime. The signal from a merger of  massive black hole binary  is bright enough to allow for detection of small violation of General Relativity (GR). Across all three phases of the event, the long lasting (months) in-spiral, the merger itself, and the following ring down phase, the amplitude and the phase of the signal are set by the non linear structure of GR. Even small deviations from GR will cause detectable phase  shifts during the in-spiral phase. After the extremely powerful burst of waves during the merger phase, the final single BH dissipate it asymmetries, and relaxes to an axi-symmetric stationary BH,  via  the emission of  radiation from its of quasi-normal mode vibrations. The frequency and damping time of these modes are only set, for a Kerr BH, by mass and spin. Violations of this prediction, will allow to discriminate between a GR, Kerr BH and any other possible exotic object.

Finally BH binary mergers are considered Ôstandard sirensÕ, as the absolute luminosity is
written  in the time derivative of the frequency, thus giving a direct, independent way of measuring the luminosity distance $D_{\ell}$ to the source, an extraordinary cosmographic tool. When associated with and independent measurement of the source red-shift, this ability would give completely independent data points in the   $D_{\ell} \left( z \right)$ curve. The red-shift of the source cannot be derived from the GW signal, being entangled with the BH mass values. Thus it needs to be measured from the electromagnetic emission of the same source. Mergers at redshift z ² 1 Ð 2 are so intense that  eLISA may locate  them to within 1-10 square degrees, sufficient to allow for searching for electromagnetic counterparts  using future surveys such as LSST. With a measurement of z , eLISA observations will allow $1 \% $ measurements of $D_{\ell} \left( z \right)$ for the majority of the sources,  completely independent all existing surveys, including supernovae and cosmic microwave background, and of any distance ladder. 

A similar manyfold impact is expected from the observation of EMRIs. For instance EMRIs will probe the population of  stellar-mass BH in the nuclei of Galaxies, a largely uncharted astrophysical ground. They will also give measurement of mass and spin of the central massive BH for the lightest galaxies, a piece of information complementary  to that given by the massive BH mergers.

Even more impacting is the role of  EMRIs as a laboratory  for studying gravity at, or near the event horizon of a massive BH. 

The  stellar-mass compact object object will orbit $\simeq 10^{5} $ 
time at a few  Schwarzschild radii of the central black hole. The emitted wave will trace the orbit of the falling object, and these will give a map of space-time  near the horizon.

Within  the Kerr metric, the complicated orbit structure is described by just mass and spin. The exquisite precision with which multipole moments can be measured from  an EMRI signal , will allow to detect relatively small deviation from this simple prediction. More important,  different sources of  deviation will leave different fingerprints in the signal that would make them distinguishable from each other .  Many mechanisms of deviation have been proposed. For instance, exotic, horizonless objects still compatible with GR, would be recognisable because the signal will not disappear abruptly when the compact object crosses the event horizon. And other signal features are predicted for deviation coming form violations of GR. 
 
Last but not least, EMRIs are  standard sirens too.
 
The impact on cosmology and fundamental physics of the detection of a cosmological stochastic background of  GW cannot be underrated.  Such backgrounds are predicted by many different scenarios of vacuum phase transition in the very early universe, in the epoch before recombination and the release of microwave background. Extra dimensions, brane-world scenarios, supersymmetry, cosmological strings, all produce distinctive backgrounds that might be detected by eLISA. 

On the opposite side of more classical astrophysics, eLISA vast and complete survey of galactic, often invisible ultra compact binaries, even beyond the Galactic centre, will constitute a unique probe of the formation and evolution history of the Galaxy. Furthermore, the brightest system will allow to study the dynamic of the binary itself including possible mergers.

And, again, the prediction of such a scientific impact is just the result of what we know today. In addition to this, it is reasonable to expect  some unpredicted discoveries  that an entirely new observational tool usually brings about.

\subsection{Possible improvements }
\label{sec:2.3}

Various improvement may be possible in the course of the studies that will bring to final mission design for eLISA 

For instance, it is possible that a third arm could be added  to eLISA, as this addition would basically imply only additional recurring costs and may then as well fit, after some design optimisation, into the cost and mass envelop of an ESA L mission.
The addition of a third arm, like in the original LISA configuration,  allows the identification of the polarisation of the wave without the need to wait for the satellite constellation to substantially rotate in space. This feature allows to better disentangle signal parameters during "short"  (days-weeks) events, like the final mergers of massive BH. The largest benefit is a higher resolution in the measurement of  the distance to the source and in the identification of its location, capabilities that  both  would enhance the cosmographic power of these observations. For reference LISA, a three-arm 5 million kilometre size constellation,  would be capable of localising the source direction to better than 1 square degree before the  merger and for 50\% of the sources at z=1, it would improve to under 3 arcmin with post-analysis after the merger. On these same signals, LISA would get  uncertainties of less than 1$\%$ in the luminosity distance for black holes at z = 1 with component masses in the range $10^5 M_ {\odot} < M(1+z) < 5 \times10^6 M_ {\odot}$ . LISA would also have the capability to measure distance to EMRI with $SNR > 50$ to $3\%$ or better with a sky position better than 5 square degrees.

A third arm, allowing for signal combinations that are insensitive to GW,  would also increase the accuracy of the estimate of the instrument  noise, thus enhancing the capability of discriminating a possible GW background.

Within the definition phase of the mission, the  possibility of restoring a third arm  will certainly be seriously considered.

\section{LISA Pathfinder}
\label{sec:3}
LISA Pathfinder \cite{LPF} is an ESA mission devoted to the verification of the instrument concept and technology for LISA-like GW missions such as eLISA.  By verification we mean here a demonstration that the hardware designed for eLISA can  reach the performance required by  the observing power described in \ref{sec:2.1}. 
The main factors setting the performance of eLISA can be discussed with reference to one of the links that constitute an arm. At the receiver satellite the frequency $\nu$ of the laser beam appears varying in time  according to \cite{Congedo}:

\begin{equation}
\label{eq:1}
 \dot{\nu }_ {r} = \dot{\nu }_ {e}+\nu _o  \left( \dot{h}_{e}-\dot{h}_{r} \right)+\frac{\nu _o}{c} \left(a_{S_{e}}-a_{S_{r}} \right)
\end{equation}

Here  quantities marked with e have been measured at the emitting satellite \emph{at the event of emission}  and those marked with r have been measured at the receiving satellite \emph{at the event of reception}. In equation \ref{eq:1} h is the metric perturbation due to the GW, $\nu_o$ is the unperturbed frequency of the beam and $a_{S}$ is the acceleration of the satellite (either receiver or emitter) along the direction of the laser beam, relative to the satellite local inertial frame. 

Equation \ref{eq:1} shows the first two sources of disturbance: the accelerations of the satellites due to forces acting on them, and  the frequency noise in the laser  $\nu _ {n}$  that corrupts  $\dot{\nu }_ {e}$. To these, one must add the  noise in the opto-electronics that perform  the local measurement of $ \dot{\nu }_ {r}$ by $ \dot{\nu }_ {loc}$ .

Furthermore, as discussed in section \ref{sec:1}, satellites accelerations are measured with respect to a reference TM,  and then subtracted from the signal in equation \ref{eq:1}. This measurement is performed by a local laser interferometer, the signal $\Delta a$ of which should nominally be given by $\Delta a=a_{S} - a_{TM}$. Here $a_{TM}$ is the acceleration of the TM relative to the local inertial frame along the direction of the link . In reality some  interferometer noise will add to the signal, appearing as an equivalent acceleration $a_{n}$ noise. In addition, because of misalignments and offsets between the reference frames relative to which interferometric measurements are performed,  the motion of other degrees of freedom of both the TM and the satellite,  contaminate the measurement with an equivalent acceleration $a_{ref}$,  that depends then on angular velocities and accelerations, and centre of mass linear accelerations, of both the TM and the  satellite.

Putting all this together we get that, after correcting for the satellite acceleration at both ends of the link, the signal at the receiver satellite becomes:

\begin{equation}
\label{eq:2}
\begin{split}
g\equiv \frac{c}{ \nu _o}\dot{\nu }_ {r} =\frac{c}{ \nu _o}\left( \dot{\nu _ {n}} +\dot{\nu }_ {loc} \right)+c  \left( \dot{h}_{e}-\dot{h}_{r} \right)+\\
+ a_{TM_{e}}-a_{TM_{r}} +a_{n_{e}}-a_{n_{r}}+a_{ref_{e}}-a_{ref_{r}}
\end{split}
\end{equation}

Of all the disturbances appearing in equation \ref{eq:2}, only the frequency noise is propagated throughout the entire constellation. Al other source of noise are local properties of the various satellites. Thus for instance, no significant correlation is expected among the forces acting on TMs located one million kilometre apart, and even the reference frame noise is due to relative misalignments and accelerations of the satellite, TMs, and \emph{local} inertial frames.

Within GW interferometers, laser frequency noise is suppressed by comparing the frequency of beams that have been generated at  same place and time, but have traveled along different arms. Thus the  beams that are compared have been  generated with same frequency, and any difference found at time of measurement, can be attributed to either the motion  of  TMs or to GW. On ground, this is achieved by making the  arms of equal length, within an  accuracy sufficient to guarantee that the light collected back at the centre from both arms, has  been emitted with negligible delay. In space the same results are obtained with a technique called Time-Delay Interferometry (TDI) described in \cite{TDI}.

The verification of the instrument concept and technology for  eLISA does not require a flight test for all the noise sources above. For instance TDI can be reasonably reproduced on ground, and the critical parts tested in the laboratory. Suppression of the frequency noise by fast phase-meters, with performances matching those required by eLISA, have indeed been demonstrated in different laboratories \cite{TDI_JPL} \cite{TDI_Fl}. Once corrected by TDI, residual laser frequency noise is calculated to contribute to total noise budget by $\approx 5\%$ above some  $\approx 8 mHz$, and to be negligible at lower frequencies. Similarly, local interferometer noise, that contributes by  $\approx 65\%$ in the same $\gtrsim 8 mHz$ range,  and is also negligible at the lower frequencies, can be satisfactorily tested in the laboratory.  

The need for a flight test originates from force noise and from reference frame noise, that contribute $100\%$ of noise at the lowest frequencies, and for which a fully representative test can only be performed in 0-g. Fortunately both  force and reference frame noise are locally generated, and thus \emph{can be fully tested within an experiment contained in just one satellite}.  

\begin{figure}

\includegraphics[trim= 0cm 6cm 0cm 4cm, clip=true, width=0.5\textwidth]{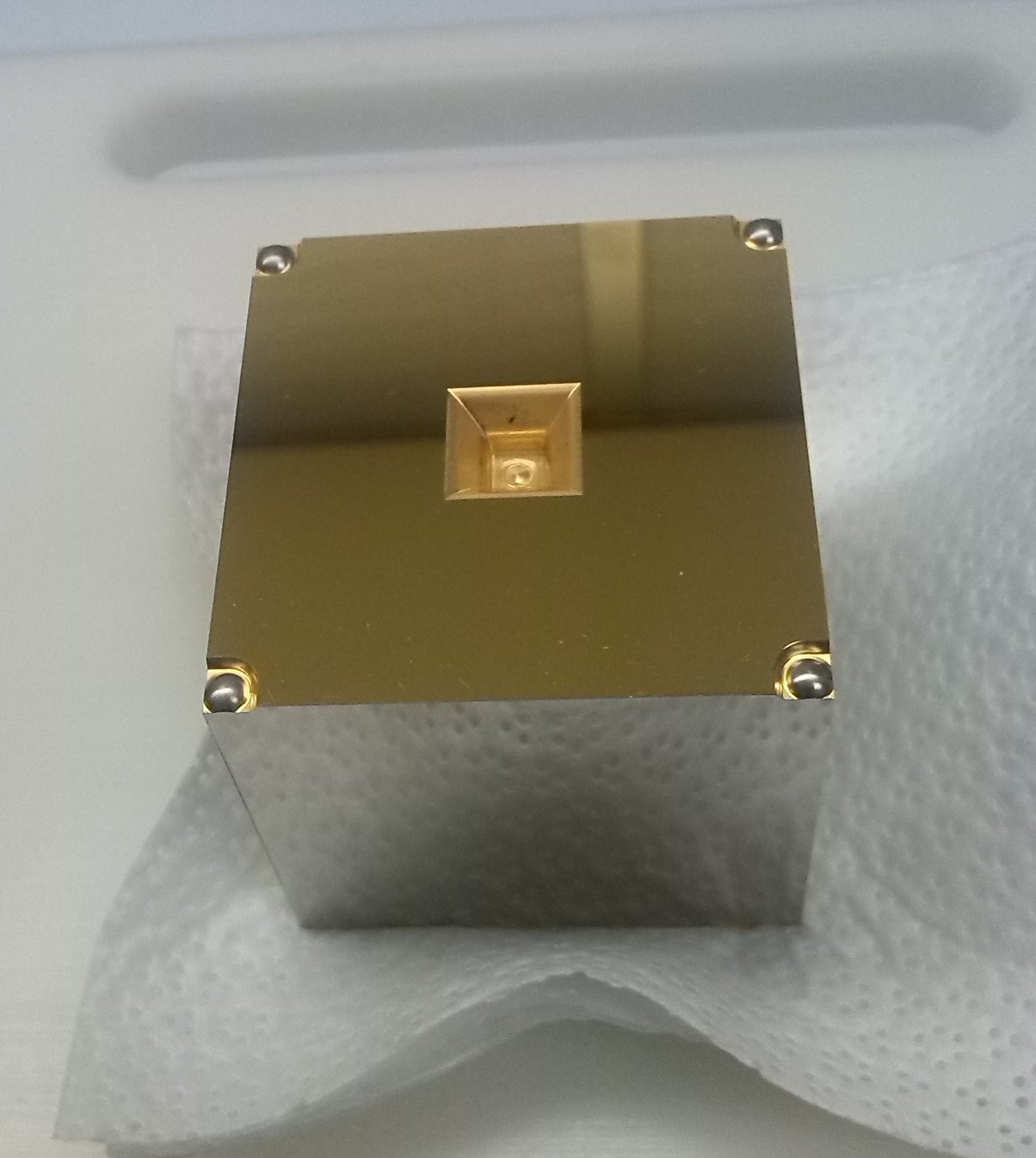}
\includegraphics[width=0.5\textwidth]{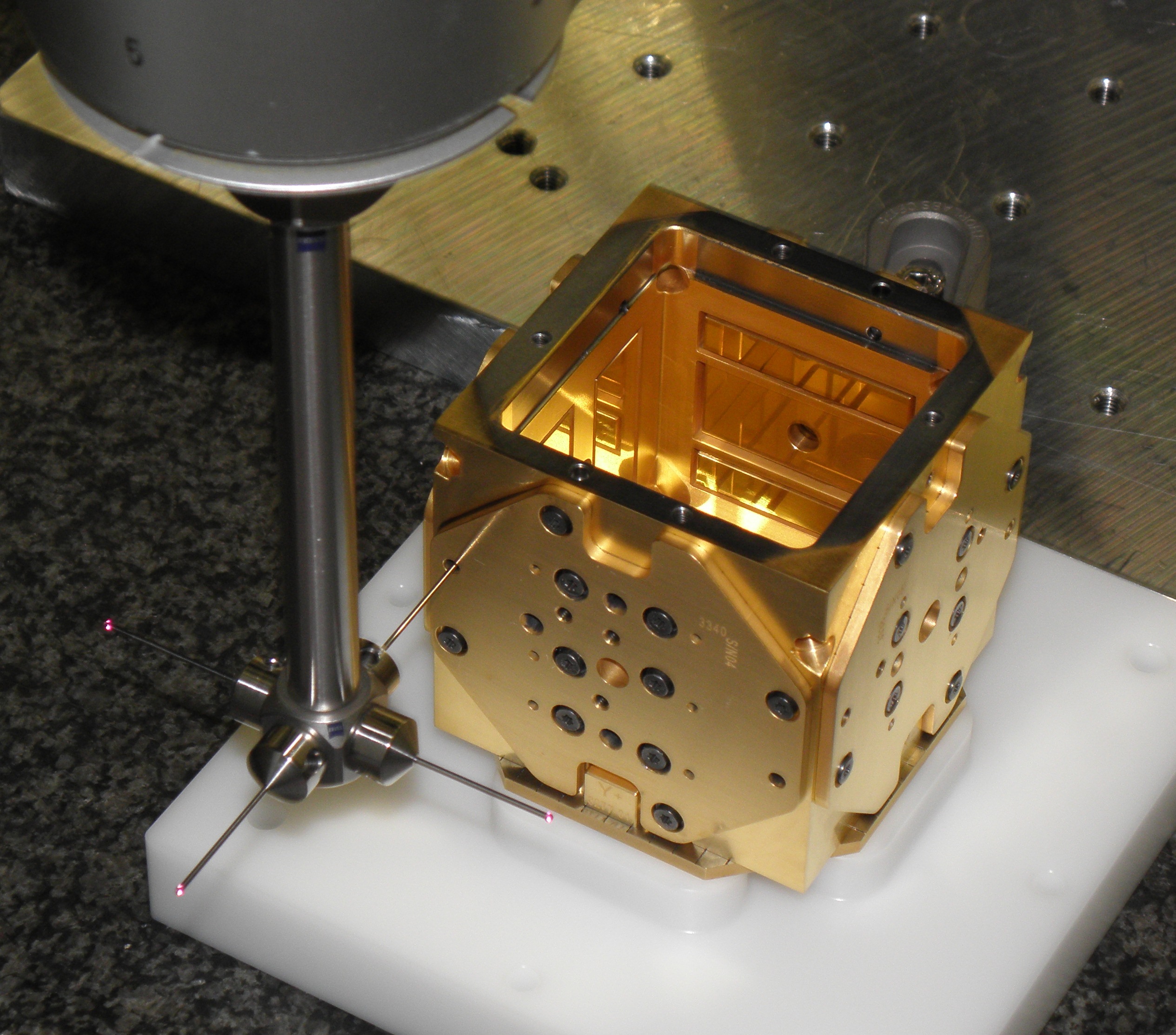}

\caption{Left: one of the the Au-PT TMs. The features in the corners and that in the centre of the top face are the interface to the launch lock and release mechanisms. Right: a view of the electrode housing with one of the sides removed to show the electrodes inside\emph{(Courtesy of CGS Italy)}}

\label{fig:4}   

\end{figure}

\begin{figure}

\includegraphics[width=0.8\textwidth, center]{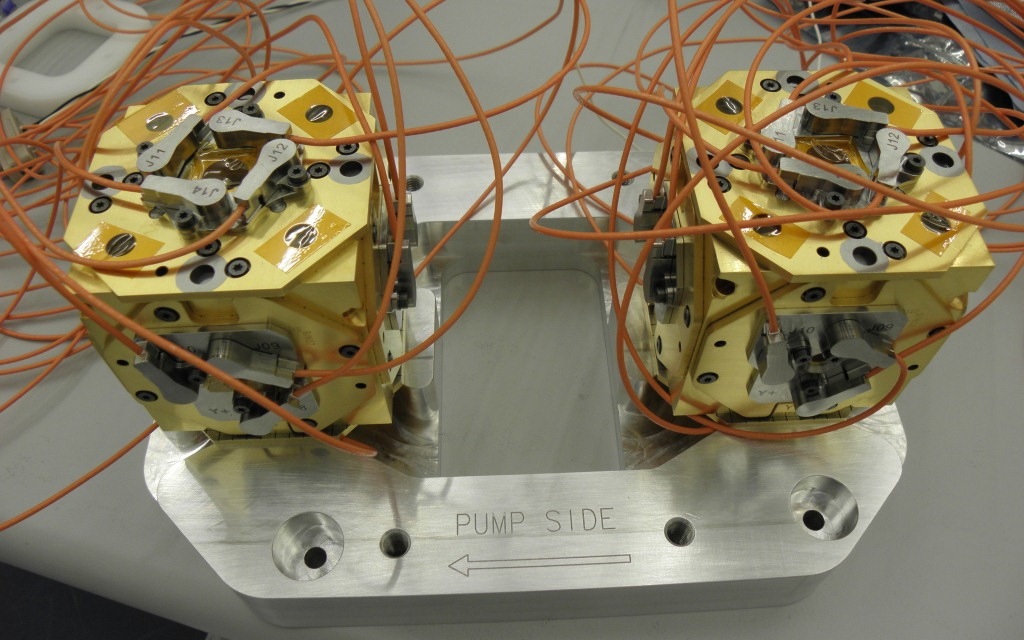}

\caption{The flight models of the electrode housings of LPF after thermal-vacuum testing\emph{(Courtesy of CGS Italy)}}

\label{fig:5}   

\end{figure}

Based on the above, the LPF concept is that of performing a full verification of the physics and technology of the \emph{local} part of eLISA instrument configuration. To implement this program ,LPF consists in essence of a miniature replica of one eLISA arm. Indeed the core payload, called the LISA Technology Package (LTP)  includes:

\begin{itemize}
\item Two free-falling TMs enclosed in their electrode housings \cite{GRS} (figures  \ref{fig:4} and \ref{fig:5}) and separated by a nominal distance of $0.38 cm$.
\item Two laser interferometers that measure the relative displacement of the two TMs and their displacement relative to the satellite, ideally along the direction x of the line joining the centres of mass of the TMs.
\end{itemize}

Each TM is equipped with all the devices that are needed for its operations and that, together, form the GRS. Besides the electrodes in the electrode housing these include: a launch lock mechanism to hold the TM in place during launch; A release mechanism to inject the TM with low momentum into its geodesic orbit; A UV-light discharging system to keep the TM neutral and compensate the electrical current due to cosmic rays; Finally a vacuum enclosure that includes the  TM, the electrode housing, the lock and release mechanisms, and the termination of the optical fibres that carry the UV light. This enclosure is needed to prepare and store the  TM in high vacuum prior to launch. 

\begin{figure}
\includegraphics[width=0.8\textwidth, center]{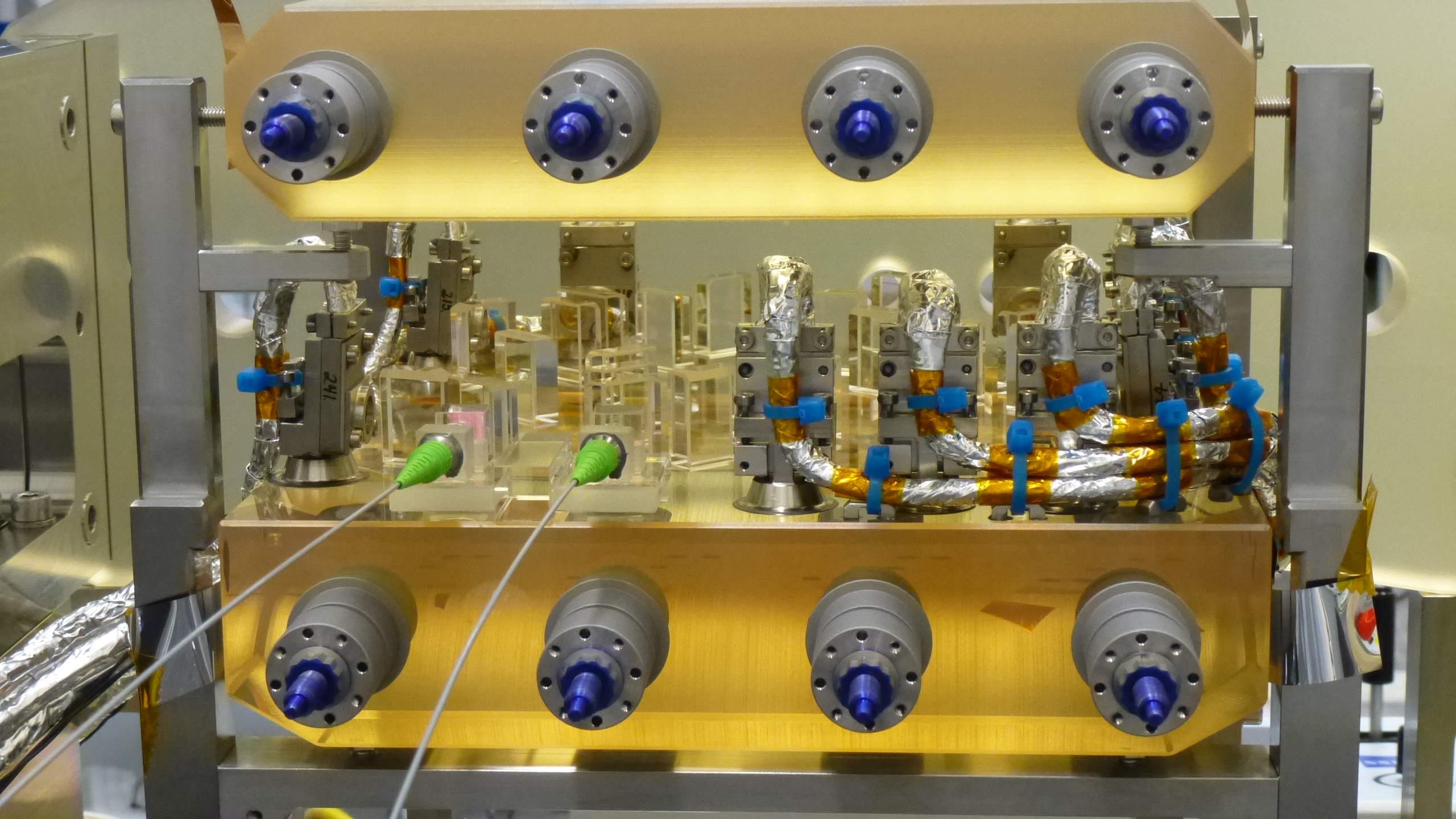}
\caption{LPF optical bench flight model (\emph{Courtesy University of Glasgow, University of Birmingham and Astrium GmbH, Friedrichshafen})}
\label{fig:6}   

\end{figure}

\begin{figure}
\includegraphics[width=0.8\textwidth, center]{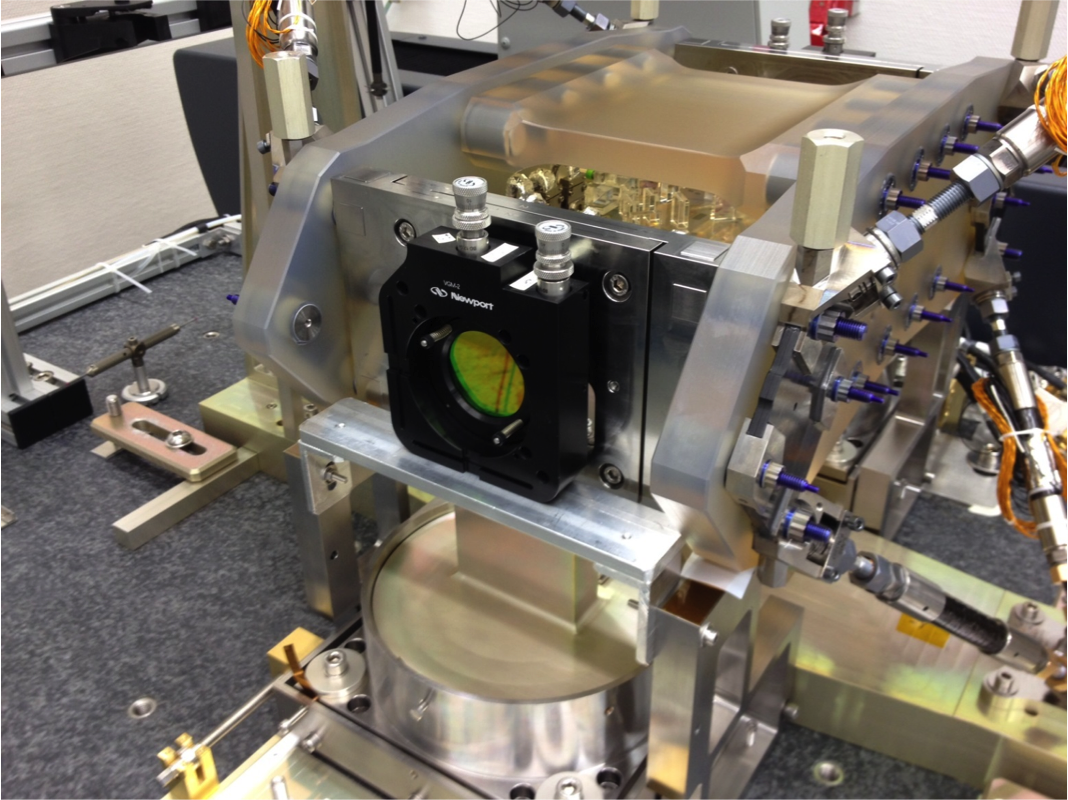}
\caption{LPF optical bench during LTP integration (\emph{ Courtesy University of Glasgow, University of Birmingham and Astrium GmbH, Friedrichshafen})}
\label{fig:7}   

\end{figure}

Both  interferometers sit on a high stability monolithic  optical  bench fabricated with the technique of silica bonding \cite{OB} (figures \ref{fig:6} and \ref{fig:7}). One interferometer is designed to measure the displacement $x_{1}$ of one TM relative to the satellite, and the other to measure the differential displacement $x_{12}$ of the TMs. Light beams are generated by a Nd:Yag NPRO 1064 nm laser and injected into the bench by fibre  injectors. Signal readout is performed by heterodyne interferometry, with frequency modulation performed via acousto-optic modulators, and  readout performed with quadrant photodiode connected to a low-noise phase-meter \cite{ifo}.

There is a difference between the basic control structure of one eLISA arm, and that of its miniature replica on LPF. In the latest, as both TMs sit in one satellite, this one cannot be forced, by the drag-free, control to follow both of them. Thus one need, to keep the system together, to apply some electrostatic control, along the arm direction, to one of the two TMs.  This is done in LPF, by forcing one of the TMs -- we call TM2-- to follow the other TM --TM1-- . Thus both the satellite, through the drag-free control, and TM2, through this so-called electrostatic suspension, are forced to follow TM1 that thus becomes the main inertial reference body for the system.

Essential to this control scheme are the $\mu N$ thrusters needed for the drag-free loop. After a  long, dedicated development of  precision ion-thrusters LPF, under schedule pressure, had finally to adopt more conventional cold-gas thrusters. These have been developed for the GAIA mission \cite{thr} and have recently been found to satisfy the requirement of LPF too.

LPF will operate on a Lissajous  orbit around the Lagrange point L1, that the satellite will reach by means of a dedicated propulsion module. The satellite and the propulsion module are ready and have successfully  undergone the full standard qualification campaign. Launch is scheduled for July 2015 .

The main goal of LPF is that of verifying that the sum of all differential parasitic forces $\Delta g$\footnote{Forces in the following are always intended per unit mass. To stress this, we prefer to indicate forces by using the symbol \emph{g} instead of  the more obvious  \emph{f} like in force.} acting on two eLISA TMs,  has a Power Spectral Density (PSD) $S_{\Delta g}$ fulfilling:

\begin{equation}
\label{eq:3}
S_{\Delta g}  \leq 30\  \frac{fm\ s^{-2}}{\sqrt{Hz}} \times \left(1+\left(\frac{f}{3 mHz}\right)^{2}\right)\hspace{1 pc} for \hspace{1 pc}1\ mHz \le f \le 30\ mHz
\end{equation}

This requirement is relaxed by a factor $\sim 7$ relative to the eLISA  requirements at 1 mHz, of  $\sqrt{2}\times3\ fm\  s^{-2}/\sqrt{Hz}$. In addition there is no requirement below 1 mHz, except that noise performance must be measured down to 0.1 mHz. This relaxation has been decided to make the the test environment easier to achieve, and to shorten all on ground verifications that otherwise, to demonstrate requirements at 0.1 Hz, may become very lengthy. However no relaxation is allowed of the hardware design.

Instrumental to the verification of  the requirement on parasitic forces, is an independent demonstration  of interferometric ranging of TM  with a sensitivity of  $9\ pm/\sqrt{Hz} \times \sqrt{1+\left(f/ 3 mHz\right)^{-4}}$, a figure similar to that required by eLISA.

In LPF data analysis \cite{ltpda}, differential forces acting on the TMs are measured, through Newton's law,  from the time series $\Delta a \left[n\right]$ of the relative acceleration of TMs along the measurement axis. The acceleration is calculated from the $x_{12}\left[n\right]$ data series by numerical double time differentiation \cite{deriv}.  
However, as the electrostatic actuation loop compensates most of the total force, this "in loop" measurement would grossly underestimate the forces. Thus the next step in the analysis is that of subtracting from $\Delta a \left[n\right]$ the series   $\Delta g_{es}\left[n\right]$ of the force that has been applied by the electrostatic suspension to TM2. The  series of the force \emph{commanded} to the electrostatic actuation system is known, but the transfer function to the real applied force must be measured. This is done  in a dedicated calibration experiment wherein  some proper,  large amplitude, calibration force  is commanded on TM2, and the resulting acceleration measured. 

The  series $\Delta g\left[n\right]=\Delta a \left[n\right]-\Delta g_{es}\left[n\right]$ represent then the forces that would accelerates the TMs if they had not been controlled. The PSD of this series must then fulfil the requirement in eq. \ref{eq:3}.

As important as showing that the PSD  of  $\Delta g\left[n\right]$ fulfils  the  requirement in \ref{eq:3}, is a quantitative verification of the physical model of the disturbance forces. To reach this goal, the master plan for the LPF on-orbit experiment  includes a sequence of investigations aimed at quantitatively identify the  main sources of parasitic forces that contribute to the noise budget. In particular specific experiments or experimental campaigns will be dedicated at quantifying the following :

\begin{itemize}
\item{Noise due to coupling of TMs to the satellite via static force gradients}
\item {Reference frame noise.}
\item {Noise due to TM random charging and charge patches}
\item{Noise due to thermal effects}
\item{Noise due to magnetic field fluctuations}
\item{Noise due to electrostatic actuation}
\item{Noise due to laser radiation pressure}
\end{itemize}

As an example, let's consider the first noise source above. The satellite generates static forces with non-zero gradient $dg/dx_{1}$, and  $dg/dx_{2}$ at the position of  TM1 and TM2 respectively. These force gradients will move with the satellite and convert its noisy motion, relative to the local inertial frame, into two  time-varying forces on TM1 and TM2 . 
To establish the contribution of these forces to  $\Delta g$,  the static force gradients will be calibrated by injecting some  proper bias signal within the control loops. The bias will exaggerate the relative motion of the satellite  and the TMs and will allow to measure $dg/dx_{1}$ and  $dg/dx_{2}$ with good signal to noise ratio. Once the gradients have been measured, the force due to  satellite motion can be calculated as   $\Delta g_{S} \left[n\right]=dg/dx_{2}\times x_{12}\left[n\right]-\left(dg/dx_{1}-dg/dx_{2}\right)\times x_{1}\left[n\right]$.

Force gradients are expected to be dominated by self-gravity and by the electric field needed for electrostatic suspension. The current estimates for both TMs gives   $|dg/dx|\le 2\times 10^{-6} s^{-2}$. If this figure is confirmed in flight, and if the performance of the drag-free controller, that sets the level of jitter of the satellite, is also confirmed, the contribution of  $\Delta g_{S} \left[n\right]$ to $\Delta g\left[n\right]$ will be negligible. In such a case,the availability of the series  $\Delta g_{S} \left[n\right]$ will allow to independently check this prediction. On the other hand, suppose that,  due to some anomaly, the gradient and/or the motion of the satellite are larger than expected --  not so remote a possibility -- so that this noise contribution is not negligible anymore. In such a case the availability  $\Delta g_{S} \left[n\right]$, not only allows to quantify the effect, but also to subtract it from  $\Delta g\left[n\right]$. The residual   $\Delta g'\left[n\right]=\Delta g\left[n\right]-\Delta g_{S} \left[n\right]$ represents then the new force data series, cleaned out of the effect of satellite motion, and only moderately corrupted by the noise introduced by the subtraction. 

The other  "noise projection" campaigns listed above may differ in the details, but  share the  approach described for the satellite noise: the availability of some signal, that once calibrated through a measured transfer function, can be used as an independent measurement of one of the dominating disturbances.The signals available for this  noise projection, besides the main $x_{12} and x_{1}$ interferometer outputs, are:

\begin{itemize}
\item{Interferometric readout of  TM rotation angles around two axes normal to x;}
\item{Capacitive readout of TMs displacement and rotation, relative to the satellite, in all degrees of freedom;}
\item{Attitude of satellite, relative to the celestial frame, measured by autonomous star trackers;}
\item{High precision thermometers at various points on both TM electrode housings;}
\item{High resolution magnetometers close to the TMs;}
\end{itemize}

In addition the following actuators are available for calibrating the various transfer functions:

\begin{itemize}

\item{Electrostatic actuation on both TMs and for all degrees of freedom;}
\item{Thruster actuation on the satellite for all degrees of freedom;}
\item{Voltage bias on all electrodes surrounding the TMs;}
\item{UV light to charge and discharge both TMs;}
\item{Coils to generate magnetic field and gradients at TM positions;}
\item{Heaters to induce temperature gradients across the electrode housings;}
\end{itemize}

It is worth noticing that the possibility of subtracting, when significant, estimated disturbances from the data, has an interest that goes beyond LPF. Once successfully  tested on LPF, this procedure will also constitute the basis for data cleaning in long lasting missions like eLISA. Indeed maintaing the instrument for months at is best working point, where all disturbance are kept within requirements, might prove demanding for various reasons.  Satellite coupling, excess charge noise, excess reference frame noise are just examples of disturbances that could instead just be measured and subtracted from the data, thus avoiding the need of continuous trimming of the instrument settings.

It is also worth noticing that, as a side product of noise projection, many physical parameters of the instrument,  that can only be measured with free-falling TMs, will be returned. These include:

\begin{itemize}
\item{Static force gradient on TM1  and on TM2.}
\item{Static difference of gravitational force $\Delta g_{dc}$ between TM1 and  TM2 along x. }
\item{Calibration of $\mu N$ thrusters}
\item{Dynamic coupling of the TMs  motion along x,  to angular and linear motion of TMs and satellite along other degrees of freedom.}
\item{Spurious pick up, by the $x_{12}$ interferometer, of angular and linear motion of TMs and satellites along other degrees of freedom.}
\item{Static parasitic voltages -- charge patches-- on electrode housing.}

\end{itemize}

These parameters are very important for the verification of the physical model of an instrument like eLISA. As an example, take the measurement of  $\Delta g_{dc}$. This parameter is very important as it sets the amount of forces that the electrostatic suspension must provide to TM2 to keep it following TM1. As the noise added by the controller scales with the amplitude of this force, there is a requirement for  $\Delta g_{dc}$  of  $\Delta g_{dc} \le 10^{-9} m/s^{-2}$. Meeting this requirement needs a careful balancing of LPF self-gravity at the positions of the TMs. This balancing is obtained by calculating the gravitational field from the position and  mass of every component of LPF, and by installing a high density balance mass next to each TM, whose field is calculated to cancel that due to the rest of LPF. Mass and position of LPF components are measured  within a sophisticated gravitational control protocol applied to all hardware suppliers.  This procedure is estimated to cancel the differential force, with an estimated residual uncertainty  of $\sim 5\times10^{-10}\ m/s^{2}$. 
 $\Delta g_{dc}$ will be measured, on LPF, during the so called drift mode experiment \cite{ff}, in which the ordinary continuous electrostatic controller of TM2, is substituted by a pulsed one. Pulses are separated by hundreds of seconds in which both TMs are not subject to any actuation. The main goal of this experiment \cite{ltpda} is that of quantifying, by difference, the extra noise due to  electrostatic actuation on TM2, a feature which is not present in eLISA. 
However, during this long free motion time stretches, $\Delta g_{dc}$  will be measured from the acceleration of the TMs, with a precision of $\sim 5\times 10^{-12} m/s^{-2}$, thus providing a key verification of the gravitational control protocol. It is worth mentioning that the gravitational control protocol is a tool that is going to be used also on eLISA with substantially similar accuracy requirements. 

The investigations that compose the LPF experiment master plan are being rehearsed within an intensive plan of simulation performed in collaboration between ESA and the science team. Investigations are  run on a  mission non-linear end-to-end simulator that has been developed by Astrium GmbH. Data analysis is performed mostly in real time, with a purposely developed data analysis toolbox \cite{ltpdatool} , in order to be able, on the basis of the results of the experiments already performed, to readjust and optimise the planning of future experiments. Up to one full week  of  operations, including  team shifts, experiment re-planning etc., has been simulated up to now. The activity is steadily ramping up in preparation for the real operations.

\begin{figure}
\includegraphics[width=0.8\textwidth, center]{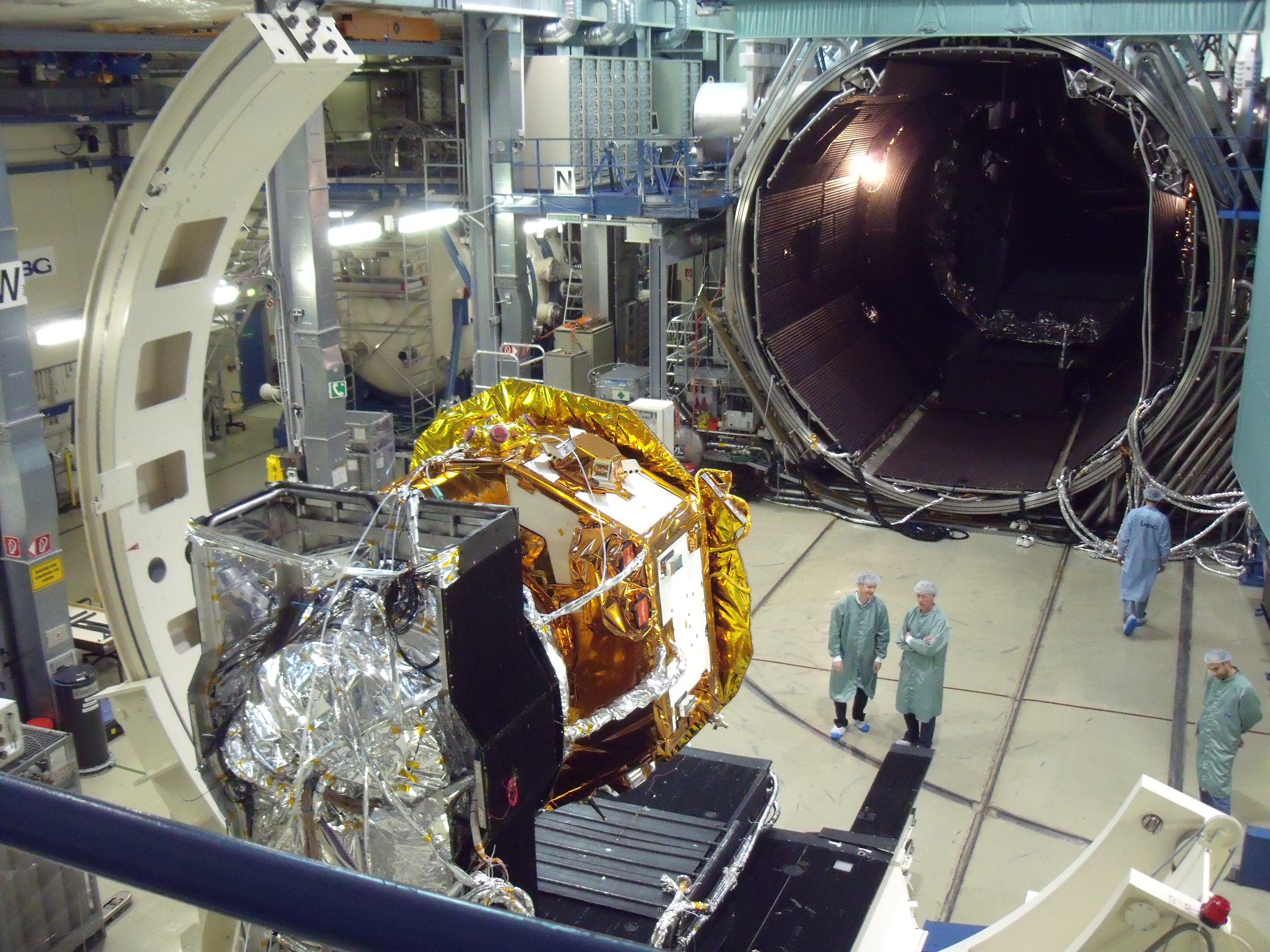}
\caption{LPF satellite ready to go into the solar simulator for the OSTT (\emph{ Courtesy of Astrium Ltd})}
\label{fig:8}   

\end{figure}

In parallel with the development and  the final delivery of the hardware, a rather rich set of tests and calibrations has allowed to consolidate the prediction of LPF in-flight performance. Many are tests at component level, but some have taken place a quite high level of integration. A remarkable example of these is a test, called On Station Thermal Test (OSTT), in which  the entire satellite  has been inserted in a solar simulator. This is  a large vacuum chamber where the satellite solar panels produce power from the light shone by a set of powerful lamps( Fig. \ref{fig:8} ). 

During this test, the flight model of the optical bench and structure ( See figures \ref{fig:6} and  \ref{fig:7}) where installed in the satellite, with the TMs substituted with steerable mirrors. This has allowed a full performance test of the interferometry in conditions representative of flight, though likely in a somewhat harsher environment. The test has shown that the requirements are met even in these conditions, that are likely to be more severe than those in flight due, for instance, to the large mechanical noise of the laboratory. 

The TMs free-flight conditions cannot be reproduced  on ground. However representative measurements of parasitic forces acting on the surface of the TMs could be performed with torsion pendulums \cite{1massa} \cite{4masse} at a quite interesting sensitivity.  Indeed a TM attached to the inertial member of a torsion pendulum may achieve a high dynamical isolation for the motion in the horizontal plane. It has been possible then to perform investigation for many of the less well known and most dreaded disturbances, including thermal gradient effects \cite{radiometrico}, brownian noise \cite{brownian} and charge patches \cite{patches}.  In addition the pendulum test bench noise provides a quite interesting upper limit for any non modelled surface force that has been overlooked in the analysis \cite{patches}. 

Putting together the end-to-end ground testing of the interferometry, the analysis and measurements of magnetic field  and of the magnetic properties of the TMs, the torsion pendulum results, and finally the simulation of system dynamics with the end-to-end simulator, a best estimate of the projected sensitivity of the LPF experiment can be derived  \cite{noise2012}. An updated summary is shown in  table \ref{tab:2} , where the projected sensitivities are also compared to LPF and eLISA requirements and with the pendulum results.

\begin{table}
\begin{center}
\caption{$\Delta g$ performance $\left[fm s^{-2} /\sqrt{Hz}\right]$ }
\label{tab:2}      
\scalebox{1}{\begin{tabular}{l c l c l c l c l c l c l c l c |}
\text{f[mHz]} & 0.10 & 0.50 & 1.0 & 5.0 & 10. & 30. \\
\hline\noalign{\smallskip}
 \text{Pendulum} & 1200. & 83. & 49. & 39. & 100. & 970. \\
 \text{LPF requirements} & 30. & 31. & 33. & 110. & 360. & 3000. \\
 \text{LPF estimate (controlled)} & 27. & 12. & 7.9 & 8.4 & 23. & 210. \\
 \text{LPF estimate(drift mode)} & 13. & 4.3 & 3.0 & 7.1 & 24. & 210. \\
 \text{eLISA requirements} & 7.1 & 6.0 & 5.8 & 13. & 48. & 430. \\
\end{tabular}
}
\end{center}
\end{table}

As it can be seen in the table \ref{tab:2}, at some frequencies LPF is expected to verify geodesic motion at or slightly better then eLISA requirements. In addition, pendulum measurements  put an upper limit to non-model force disturbances at, or even slightly better than LPF requirements across all the  target bandwidth. 

\section{Concluding remarks}
\label{sec:4}

The data in table \ref{tab:2}  suggest that  a sensible approach toward an orbiting GW observatory like eLISA is being followed. In this approach the risks connected with an entirely new space instrumentation are retired in two steps --ground testing $\to$ LPF $\to$ eLISA-- each one involving a leap in performance, relative  the previously demonstrated level, of less than one order of magnitude. In this two-step approach  a crucial milestone  is the launch and operation of LPF. 

With the current planning, LPF operations, including data analysis, will be completed by 2016.  A few years later, advanced LIGO and Virgo are expected to start observations, with useful SNR, of stellar mass binaries, at least at a monthly rate, and quite plausibly even multiple events per week when the detectors reach their most sensitive configurations. It is also possible  that pulsar timing may do some observations, at low but significant SNR, within the same time frame.

If this prediction is maintained, the next 5 years will witness a major revolution in the field of GW wave astronomy, with the completion of  the era of technological development, and the transition to that of astronomical observation, with the implementation and/or the operation of working observatories of increasingly deep reach.

\begin{acknowledgements}

I thank   Karsten Danzmann and Paul McNamara  for their critical reading of the manuscript. 
\end{acknowledgements}

\end{document}